\documentclass{aa}
\usepackage{txfonts}
\usepackage{units}
\usepackage{booktabs}
\usepackage{graphicx}
\usepackage{natbib}
\bibpunct{(}{)}{;}{a}{}{,}

\title{The baryon density at $z=0.9-1.9$}
\subtitle{Tracing the warm-hot intergalactic medium with broad Lyman $\alpha$ absorption}

\author{N.\ Prause \and D.\ Reimers \and C.\ Fechner \and E.\ Janknecht}
\institute{Hamburger Sternwarte}

\date{Received 12 February 2007/ Accepted 09 May 2007}

\abstract {}{
We present an analysis of the Lyman $\alpha$ forests of five quasar spectra in the near UV. Properties of the intergalactic medium (IGM) at an intermediate redshift interval ($0.9\leq z\leq1.9$) are studied. The amount of baryons in the diffuse photoionised IGM and the warm-hot intergalactic medium (WHIM) are traced to get constraints on the redshift evolution of the different phases of the intergalactic gas.}{
The baryon density of the diffuse IGM is determined with photoionisation calculations under the assumption of local hydrostatic equilibrium. We assume that the gas is ionised by a metagalactic background radiation with a Haardt \& Madau (2001) spectrum. The WHIM is traced with broad Lyman $\alpha$ (BLA) absorption. The properties of a number of BLA detections are studied. Under the assumption of collisional ionisation equilibrium a lower limit to the baryon density could be estimated.}{
It is found that the diffuse photoionised IGM contains at least $\sim25\%$ of the total baryonic matter at redshifts $1\lesssim z\lesssim2$. For the WHIM a lower limit of $\sim2.4\%$ could be determined. Furthermore the data indicates that the intergalactic gas is in a state of evolution at $z\sim1.5$. We confirm that a considerable part of the WHIM is created between $z=1$ and $z=2$.}{}


\begin{document}
\maketitle
\section{Introduction}
The baryonic matter in the universe can be divided into four gas phases. The largest fraction of baryons resides in the diffuse IGM which is photoionised by a metagalactic UV background radiation. While at higher redshifts $z\gtrsim3$ about $90\%$ of the baryonic matter is assembled in the photoionised IGM \citep{Weinberg1997}, in the local universe this gas phase inherits a much smaller fraction of matter ($\sim30\%$)  \citep{Penton2004, Lehner2007}. This incident is called the missing baryon problem. It is now widely believed that the gas has been shock-heated by gravitational collapse to higher temperatures $T\sim\unit[10^5-10^7]{K}$ thus forming the WHIM \citep{Valageas2002}. It is expected to contribute a major fraction of the baryonic matter at low redshifts \citep{Cen1999, Dave2001, Dave2003}. Hot gas that is gravitationally bound in galaxy clusters as well as condensed objects like stars and cool galactic gas only contain about $10\%$ of the total baryons each \citep{Dave2003, Bell2003}.

The WHIM is a low density ($n_{\rm{H}}\sim\unit[10^{-6}-10^{-4}]{cm^{-3}}$) high temperature plasma and is thus hard to detect. Current X-Ray observatories are not sensitive enough to detect its diffuse emission which is suspected to have a very low surface brightness. However, the analysis of quasar absorption features seems to be a promising approach.

Oxygen is a favoured element since it is relatively abundant and its transitions have large oscillator strengths. The \ion{O}{vi} ion traces gas at a temperature of $T\sim\unit[3\cdot10^5]{K}$, characteristic for the WHIM. At low redshifts ($z<0.5$) various detections of \ion{O}{vi} absorbers that are associated with the WHIM have been reported \citep{Tripp2000, Oegerle2000, Tripp2000b, Chen2000, Savage2002, Richter2004, Sembach2004, Savage2005, Tripp2005, Danforth2005}. These measurements imply a baryon density of the WHIM in \ion{O}{vi} absorption of $\Omega_{\rm{b}}(\ion{O}{vi})\geq0.0022h_{\rm70}^{-1}$, assuming that 20 percent of the oxygen is fivefold ionised and that the mean oxygen abundance of the IGM is 0.1 solar. This relatively low value is explained with the absence of \ion{O}{vi} in hotter gas with temperatures $T\geq\unit[10^6]{K}$ which might contain a large fraction of baryons. Higher oxygen transition (\ion{O}{vii} and \ion{O}{viii}) could be detected with X-ray absorption measurements, however, the low spectral resolution of current X-ray observatories limits the possibilities in this research field. 

Currently the search for broad Lyman $\alpha$ (BLA) absorption of neutral hydrogen seems to be the most promising approach. Due to the high gas temperature, WHIM absorption lines will be strongly broadened and due to the expected small neutral fraction ($f_{\rm{HI}}\lesssim10^{-5}$) most lines will be shallow. Some detections of BLAs at low redshifts have been reported in the last years \citep{Tripp2001, Richter2004, Sembach2004, Richter2006, Lehner2007}. \citet{Richter2006} derived a lower limit of $\Omega_{\rm{b}}(\rm{BLA})=0.0027h_{70}^{-1}$ for the WHIM baryon density. It is expected that the WHIM makes up at least $20\%$ of the baryons in the local universe \citep{Lehner2007}.

In this paper we estimate the amount of baryons residing in the IGM, and the WHIM in particular, at an intermediate redshift interval ($0.9\leq z\leq1.9$). The evolution of the gas phases is of particular interest. In Sect.\ \ref{sect:properties} some properties of the IGM are discussed. Sect.\ \ref{sect:observations} describes the data used in this work. BLA detections in the sightline of five quasars are reported and discussed in Sect.\ \ref{sect:results}. A summary of the results will be given in Sect.\ \ref{sect:discussion}. 

\section{Properties of the IGM}
\label{sect:properties}
\subsection{The baryon density}
To make an estimation of the baryon content of the IGM, the total hydrogen density has to be known. Since the analysis of quasar absorption lines just gives the column density of neutral hydrogen and the gas is highly ionised, the ionisation fraction is an important factor for an estimation of the baryon density. With knowledge of the total hydrogen column density $N_{\rm{H}}$ of each absorber, the density parameter $\Omega$ can be summed up:
\begin{equation}
\label{eq:density}
\Omega_{\rm{b}}=\frac{\mu m_{\rm{H}}H_0}{\rho_cc}\sum_{ij}N_{{\rm H},ij}/\sum_j\Delta X_j,
\end{equation}
where $\mu=1.3$ is a correction factor for the helium abundance, $m_H$ the hydrogen mass, ${H_0}$ the Hubble constant, $\rho_c$ the critical density, $c$ the speed of light and $\Delta X$ is the comoving path length (Eq.\ \ref{eq:path}). The index $i$ denotes a Lyman $\alpha$ system along the line of sight $j$.

The comoving path length $\Delta X$ was calculated using a $\Lambda$CDM cosmology:
\begin{equation}
\label{eq:path}
dX=\frac{1+z}{\sqrt{\Omega_{\rm{M}}(1+z)}+\Omega_\Lambda/(1+z)}dz,
\end{equation}
with the dark energy contribution $\Omega_\Lambda=0.73$, the dark matter density $\Omega_{\rm{DM}}=0.23$ and the baryonic matter density $\Omega_{\rm{BM}}=0.04$ \citep{Spergel2003}.

Since the diffuse photoionised IGM and the WHIM are governed by different physics, different approaches have to be made to determine the ionisation fraction.

\subsection{Photoionised IGM}
The ionisation fraction of the diffuse photoionised IGM is mainly dependent on the photoionisation rate of the ionising background radiation, the gas temperature and the gas density. In this section some properties of the photoionised IGM are discussed that will be used to estimate the density parameter. 

The neutral fraction of a highly ionised optically thin gas can be written as
\begin{equation}
\label{eq:neutral}
\frac{n_{\ion{H}{i}}}{n_{\rm{H}}}=n_{\rm{e}}\alpha_{\ion{H}{ii}}\Gamma^{-1}\sim0.46n_{\rm{H}}\Gamma_{12}^{-1},
\end{equation}
where $\alpha_{\ion{H}{ii}}\approx\unit[4\cdot10^{-13}T^{-0.76}]{cm^3s^{-1}}$ is the recombination coefficient of ionised hydrogen, $T$ is the gas temperature and $\Gamma=\unit[\Gamma_{12}\cdot10^{12}]{s^{-1}}$ is the photoionisation rate of the UV background radiation.

Under the assumption of local hydrostatic equilibrium, the typical absorber size should be of the same order as the local Jeans length. In this case the total hydrogen column density $N_{\rm{H,J}}$ can be expressed in terms of the total hydrogen volume density $n_{\rm{H}}$ and the gas temperature $T=\unit[T_4\cdot10^4]{K}$ \citep{Schaye2001}:
\begin{equation}
\label{eq:jeans}
N_{\rm{H,J}}\sim\unit[1.6\cdot10^{21}]{cm^{-2}}n_{\rm{H}}^{1/2}T_4^{1/2}\left( \frac{f_g}{0.16}\right),
\end{equation}
where $f_g$ is the fraction of mass in gas. On the scales of interest $f_g$ will not be far from its universal value $f_g=\frac{\Omega_b}{\Omega_M}\approx0.16$, however in cold collapsed clumps it could be close to unity.

Substituting the neutral fraction (Eq.\ \ref{eq:neutral}) into Eq.\ \ref{eq:jeans} gives the neutral hydrogen column density $N_{\ion{H}{i}}$:
\begin{equation}
\label{eq:jeans2}
N_{\ion{H}{i}}\sim\unit[2.3\cdot10^{13}]{cm^{-2}}\left(\frac{n_{\rm{H}}}{\unit[10^{-5}]{cm^{-3}}}\right)^{3/2}T_4^{-0.26}\Gamma_{12}^{-1}\left( \frac{f_g}{0.16}\right).
\end{equation}

The photoionisation rate $\Gamma$ can be integrated with knowledge of the UV background intensity $J(\nu)$ and the photoionisation cross section $\alpha(\nu)$:
\begin{equation}
\label{eq:photo}
\Gamma=4\pi\int_{\nu_0}^\infty\frac{J(\nu)}{h\nu}\alpha(\nu)d\nu,
\end{equation}
where $h$ is the Planck constant and $\nu_0$ is the frequency at the hydrogen ionisation energy.

\subsection{WHIM}
The WHIM is a low density, high temperature plasma and is thus expected to create broad and shallow absorption features in neutral hydrogen. The ionisation fraction can be estimated, following a work by \citet{Sutherland1993}. They have investigated cooling functions for low-density astrophysical plasmas under the assumption of collisional ionisation equilibrium. For the temperature range of interest ($T\sim\unit[10^5-10^7]{K}$) the ionisation fraction $f_{\rm{ion}}=\frac{N_{\ion{H}{i}}+N_{\ion{H}{ii}}}{N_{\ion{H}{i}}}\approx\frac{N_{\ion{H}{ii}}}{N_{\ion{H}{i}}}$ can be approximated as a polynomial \citep{Richter2006}:
\begin{equation}
\label{eq:ion}
\log f_{\rm{ion}}(T)\approx-13.9+5.4\log T-0.33(\log T)^2.
\end{equation}
It is solely dependent on the gas temperature $T$ and does not vary over a wide range of metallicities. However, it is just valid for pure collisional ionisation equilibrium. \citet{Richter2006b} found that photoionisation is important for the WHIM as well and that neglecting it might underestimate the baryon density up to $50\%$. The best fit to their simulated data is a linear relation between the ionisation fraction and the gas temperature:
\begin{equation}
\label{eq:ion2}
\log f_{\rm{ion}}(T)\approx-0.75+1.25\log T.
\end{equation}

The temperature of an absorber can be estimated from the measured Doppler parameter. The broadening of the line shape is created by a combination of thermal and non-thermal effects: $b=\sqrt{b_{\rm{th}}^2+b_{\rm{non-th}}^2}$.
For pure thermal broadening the Doppler parameter $b$ would be directly connected with the gas temperature $T$:
\begin{equation}
\label{eq:temp}
b_{\rm{th}}=\sqrt{\frac{2kT}{m}},
\end{equation}
where $k$ is the Boltzmann constant and $m$ is the particle mass. WHIM \ion{H}{i} absorbers with temperatures $T\geq\unit[10^5]{K}$ will thus create absorption features with Doppler parameters $b\gtrsim\unit[40]{km\,s^{-1}}$.

A determination of the gas temperature requires a detailed study of the line shape of each absorber. According to \citet{Richter2006} non-thermal broadening effects should be detectable in high resolution spectra with a good signal to noise ratio ($S/N$). The most important non-thermal broadening mechanisms are macroscopic velocities and turbulence within the gas. These should, in general, produce non-Gaussian line shapes. Though there could be cases in which macroscopic velocities create a Gaussian shaped broadening, e.g. a symmetric infall of gas towards the line of sight, these should not be dominant. Since WHIM absorbers will have Doppler parameters of at least $b\geq\unit[40]{km\,s^{-1}}$ the Hubble flow, which is suspected to contribute with $\sim\unit[7]{km\,s^{-1}}$ per hundred kpc absorber width, should not be dominant for typical absorber sizes.

Note that the ionisation fraction increases with the gas temperature. Consequently absorbers with higher temperatures would inhabit more baryons than colder absorbers with the same column density. Therefore systems that inherit the highest amount of baryons are the most broadened and might thus be the hardest to detect.

\section{Observations}
\label{sect:observations}
The Lyman $\alpha$ forests of five quasars were analysed in the near UV. A wavelength range from \unit[2300]{\AA} to \unit[3500]{\AA} was covered. Up to the atmospheric cut-off at $\sim\unit[3000]{\AA}$ the data was taken with the Space Telescope Imaging Spectrograph (STIS) on board the Hubble Space Telescope. Higher wavelengths were covered with the UV Echelle Spectrograph (UVES) at the ESO Very Large Telescope. For \object{HE 0515-4414} \citep{Reimers1998} data were available from both instruments thus covering a wide wavelength range. The E230M grating of STIS gives a resolution of $R\approx30\,000$. The UVES data were taken with a slit width of 0.8\arcsec\ resulting in a resolution of $R\approx50\,000$. The spectra of three additional quasars from the Hamburg ESO survey were used, that have been obtained in the ESO service mode for previous studies. \object{HE 0141-3932} and \object{HE 2225-2258} were observed with a slit width of 1.0\arcsec\ resulting in a resolution of $R\approx40\,000$ \citep{Wisotzki2000} and \object{HE 0429-4901} again with a slit width of 0.8\arcsec. Furthermore data from the STIS archive was used. The data for \object{PG 1634+706} \citep{Schmidt1983} were taken in two programs (S.\ Burles 7292/7 and B.\ Januzi 8312/8) also using the E230M grating.

Except for HE 0515-4414, pipeline data was used for all quasars. \citet{Janknecht2006b} have reworked the reduction of HE 0515-4414 and have done vacuum and barycentric corrections to the wavelength scale as well as a co-addition of multiple exposures for all quasars. Table \ref{tab:quasars} shows the redshift $z$, the signal to noise ratio $S/N$ per pixel, the used wavelength range $\Delta\lambda$ with the corresponding redshift range $\Delta z$ and the effective redshift range for the search for BLAs $\Delta z_{\rm{BLA}}$.

\begin{table}
\begin{minipage}[t]{\columnwidth}
\label{tab:quasars}
\centering
\renewcommand{\footnoterule}{}
\caption{Quasar overview}
\begin{tabular}{cccccc}
\toprule
QSO & $z$ & $S/N$ & $\Delta\lambda$ [\unit{\AA}] & $\Delta z$ & $\Delta z_{\rm{BLA}}$\footnote{Just the regions with $\left\langle S/N\right\rangle \geq15$ that are not too much populated with Lyman $\alpha$ forest lines could be used.} \\ \midrule
HE 0515-4414 & 1.73 & 10-50 & 2310-3260 & 0.782 & 0.240 \\  
HE 0141-3932 & 1.80 & $\sim$ 25 & 3061-3384 & 0.266 & 0.101\\ 
HE 2225-2258 & 1.89 & $\sim$ 25 & 3057-3478 & 0.346 & 0.033\\ 
HE 0429-4901 & 1.94 & $\sim$ 8 & 3188-3538 & 0.288 &  0.126\\ 
PG 1634+706 & 1.34 & 5-50 & 2310-2790 & 0.395 & 0.255 \\ \bottomrule
\end{tabular}
\end{minipage}
\end{table}

A complete list of all Lyman $\alpha$ forest absorbers was taken from \citet{Janknecht2006}. It was used for the photoionisation calculations and as a basis for the search for BLA absorbers. The $S/N$ varies strongly over the spectrum of each quasar. Since the possibility of detecting broad and shallow features is strongly dependent on the noise, not the whole available wavelength range was usable for the search for BLAs. An average signal to noise ratio of $\left\langle S/N\right\rangle \geq15$ was used as a lower limit to the data quality. Furthermore a region of $\Delta v=\unit[5\,000 ]{km\, s^{-1}}$ around the Lyman $\alpha$ emission line of each quasar was excluded to avoid influences of the proximity effect. This resulted in a usable redshift range of $\Delta z_{\rm{BLA}}=0.723$. The STIS and the UVES data cover nearly the same range of analysed data.

\section{Results}
\label{sect:results}
\subsection{Photoionisation Calculations}
We used  \citet{Haardt2001} spectra (HM) to calculate the photoionisation rates $\Gamma$ at different redshifts (Eq.\ \ref{eq:photo}). For the mean intensity of the ionising background at the hydrogen ionisation edge the values provided by HM were used. They increase from $\log J_\nu=-21.61$ for $z\approx0.9$ to $\log J_\nu=-21.29$ for $z\approx1.9$. The fraction of mass in gas was set to its universal value $f_g=0.16$.

The gas temperature is supposed to be about $T\sim\unit[10^4]{K}$, however, a more precise determination is very difficult. Therefore, in the following considerations the gas temperature will be left variable. Using Eq.\ \ref{eq:jeans2}, the neutral hydrogen volume density can be approximated for each absorber. The resulting total hydrogen column densities (Eq.\ \ref{eq:jeans}) can be summed up to a total baryon density using Eq.\ \ref{eq:density}. Since \citet{Schaye2001} restricts his analysis to optically thin clouds, only absorbers with column densities $\log N_\ion{H}{I}\leq17$ were used. A relatively low significance level of $SL=\frac{W}{\sigma_W}\geq1$, where $W$ is the observed equivalent width and $\sigma_W$ its $1\sigma$ error, was used for the selection of Lyman $\alpha$ lines \citep{Janknecht2006b}. The resulting sample contains 624 Lyman $\alpha$ absorbers in the lines of sight of five quasars, covering a total redshift interval of $\Delta z=2.08$, corresponding to a comoving path length of $\Delta X=5.465$.

The complete sample results in a baryon density of $\Omega_{\rm b}({\rm Ly}\alpha)=0.010\pm0.001\cdot T_4^{0.59}$. Dividing the sample into two redshift intervals shows an observable increase of the baryon content with redshift. While for $0.9\leq z<1.5$ the density parameter is found to be $\Omega_{\rm b}({\rm Ly}\alpha)=0.008\pm0.001\cdot T_4^{0.59}$, at higher redshifts $1.5\geq z\geq1.9$ it increases to $\Omega_{\rm b}({\rm Ly}\alpha)=0.012\pm0.001\cdot T_4^{0.59}$. The errors are propagated from the errors of the column density fits. The real uncertainty would be much higher. It should be remembered that the absolute value of the baryon density depends directly on the mean intensity $J_\nu$ of the ionising background, which is not well known. A change in the photoionisation rate $\Gamma$ of 20\% could cause a change in the derived baryon density of about 10\% \citep{Lehner2007}. However, the relative numbers which trace a possible variation of the baryon density would not be sensitive to this error source. Note that the results here are just applicable for a constant gas temperature. Since the temperature of the photoionised gas is expected to decrease with decreasing redshift \citep{Ricotti2000, Schaye2000} the evolution could be even stronger.

\citet{Schaye2001} proposes a more statistical approach to the problem. In the literature observational results are often expressed in terms of the number of absorption lines per unit absorption distance $X$ and per unit column density $f(N_{\ion{H}{i},z})$. Using this function, the density parameter can be integrated:
\begin{equation}
\label{eq:density2}
\Omega_{\rm{b}}=\frac{\mu m_{\rm{H}}H_0}{\rho_c c}\int_{N_{\rm{min}}}^{N_{\rm{max}}}N_{\ion{H}{i}}\frac{n_{\rm{H}}}{n_{\ion{H}{i}}}f(N_{\ion{H}{i}},z)dN_{\ion{H}{i}}.
\end{equation}
Combining this equation with equations \ref{eq:neutral} and \ref{eq:jeans2} yields
\begin{equation}
\label{eq:density3}
\Omega_{\rm{b}}\sim2.2\cdot10^{-9}h^{-1}\Gamma_{12}^{1/3}T_4^{0.59}\cdot \int_{N_{\rm{min}}}^{N_{\rm{max}}}N_{\ion{H}{i}}^{1/3}f(N_{\ion{H}{i}},z)dN_{\ion{H}{i}},
\end{equation}
where $h=\frac{H_0}{\unit[100]{km\,s^{-1}\,Mpc^{-1}}}$. The fraction of mass in gas $f_g$ was again set to its universal value $f_g=0.16$. \citet{Janknecht2006b} found that the column density distribution can be expressed as a single power law, $f(N_{\ion{H}{i}})\approx AN_{\ion{H}{i}}^{-\beta}$ with $\beta=1.60\pm0.03$ and $\log A=9.4\pm0.3$ over the total observed redshift range. Adopting an averaged photoionisation rate $\Gamma_{12}=1.07$, according to $z\approx1.5$, and the integration limits $N_{\rm min}=12.9$ and $N_{\rm max}=15.7$ \citep{Janknecht2006} results in a baryon density of $\Omega_{\rm b}({\rm Ly}\alpha)=0.009\cdot T_4^{0.59}$ which is within the $1\sigma$ error interval of the previous result. However, the density parameter strongly depends on the slope of the column density distribution $\beta$. \citet{Janknecht2006b} found a slightly different slope for the higher redshift sample $1.5\leq z\leq2$ of $\beta=1.55\pm0.04$. For this analysis the uncertainty in the determination of $\beta$ has a very big impact. Using $\beta=1.55\pm0.04$ for the high redshift sample would result in a over four times higher density parameter $\Omega_{\rm b}({\rm Ly}\alpha)=0.044\cdot T_4^{0.59}$. Due to the sensitivity of Schayes method to $\beta$, the individual treatment without using approximations for the column density distribution is to be preferred.

The unknown gas metallicity does not pose a big error source. Though the main metallicity of the IGM is suspected to be about 0.01 solar, according to \citet{Cen2005}, most of the low column density Lyman $\alpha$ absorbers should have lower metallicities ($\sim10^{-3}$ solar). We created photoionisation models, using the \textsc{Cloudy} code \citep{Ferland1998} to study the influence of metals on the baryon density. The resulting density parameter is constant for metallicities lower than 0.01 solar. We repeated the calculations with different spectra for the ionising background. Using a modified Haardt \& Madau spectrum where the helium edge is shifted to \unit[3]{Ryd}, possibly due to the opacity of higher \ion{He}{ii} Lyman series lines \citep{Fechner2006} or a background radiation mainly created by AGN as proposed by \citet{Mathews1987}, resulted in density parameters that are within each others ${1\sigma}$ error intervals. Misinterpretations in the sample of narrow Lyman $\alpha$ lines could add another uncertainty. This is expected to be small compared to the other mentioned error sources since it is only important for weak absorbers which do not contribute much to the total baryon density.

\subsection{BLA detections and WHIM baryon content}
Since the temperature and thus the baryon content of BLAs is just determinable when pure thermal broadening can be assumed, the shape of each broad absorber had to be analysed with care. To avoid too many misinterpretations an average signal to noise ratio of $\left\langle S/N\right\rangle \geq15$ per pixel was used as a lower limit to the data quality. We selected each broad absorber with a Doppler parameter of $b\gtrsim\unit[40]{km\,s^{-1}}$ by a careful by eye inspection of the spectra. Every absorber that showed signs of asymmetries in the line profile was excluded from the list. Furthermore no saturated absorbers, blended lines or features that are part of multi-component systems were used, since the real line width and shape is hard to determine in these cases. Since the continuum placement can have a large impact on the correct determination of the column density and the Doppler parameter, in particular for broad and shallow absorbers, just features were included that lie in regions with a clearly defined continuum.

All thus found absorbers were fitted with the $\chi^2$ minimising algorithm \textsc{Candalf} by R.\ Baade. The code does a Doppler profile fit to the line shape and simultaneously a polynomial fit to the continuum. The errors from the line fits are internally added to the continuum placement errors.

It has been empirically found that absorbers with $\frac{N}{b}\gtrsim\frac{2\cdot10^{12}}{\left\langle S/N\right\rangle}\unit{cm^{-2}/km\,s^{-1}}$ are detectable. Using the minimum signal to noise ratio $\left\langle S/N\right\rangle=15$ would yield a limit of $\log\frac{N}{b}:=\log\left(\frac{N}{b}\cdot\frac{\unit{km\,s^{-1}}}{\unit{cm^{-2}}}\right)\geq11.1$ for absorbers that could be detected in the whole analysed wavelength range.

The final sample was divided into two classes. All absorbers that still show slight signs of a non-Gaussian shape, multi-component substructure, line blends or whose shape is not clearly discernable because of noise were marked as tentative. At a total redshift interval of $\Delta z=0.723$, 38 broad absorbers that are possibly originated in the WHIM were found, including 29 tentative cases. Of the 9 good detections, 8 had a sensitivity coefficient of $\log\frac{N}{b}>11.1$ and would thus be detectable throughout the analysed wavelength range. With a total number of 9 (38) absorbers, the number of detections per unit redshift is $\frac{dN}{dz}=13.8\ (51.2)$. The bracketed values represent the whole sample, including the tentative detections. Table \ref{tab:BLAs} shows the column density $N_{\rm{HI}}$, the Doppler parameter $b$ and the temperature $T$ of all selected broad absorbers. The tentative detections are marked in the last column. The most reliable WHIM detections are shown in Fig.\ 1.

\begin{table}
\begin{minipage}[t]{\columnwidth}
\label{tab:BLAs}
\centering
\renewcommand{\footnoterule}{}
\caption{WHIM candidates}
\begin{tabular}{lllll}
\toprule
$z$ & $\log N_{\rm{HI}}$\footnote{$\log N_{\rm{HI}}:=\log\left(\frac{N_{\rm{HI}}}{\unit{cm^{2}}}\right)$} & $b\ [\unit{km\,s^{-1}}]$ & $T$ [K]\footnote{The temperatures were determined under the assumption of pure thermal broadening.} & tent\footnote{The tentative detections are marked: (S) $S/N$ too low, (B) line blend, (G) non-Gaussian shape, (M) multi-component substructure}\\ \midrule
\multicolumn{5}{c}{HE 0515-4414}\\ \midrule
 1.0624 & $13.39\pm0.08$ & $\ \ 51\pm9$ & $1.58\cdot10^{5}$  & S  \\ 
 1.0780 & $13.13\pm0.14$ & $\ \ 58\pm18$ & $2.07\cdot10^{5}$  & S \\ 
 1.1146 & $13.33\pm0.06$ & $\ \ 51\pm9$ & $1.59\cdot10^{5}$ & S \\ 
 1.1276 & $13.50\pm0.08$ & $101\pm17$ & $6.15\cdot10^{5}$ & B \\ 
 1.1599 & $13.44\pm0.09$ & $\ \ 42\pm8$ & $1.08\cdot10^{5}$ & SG \\ 
 1.1632 & $14.02\pm0.04$ & $\ \ 57\pm5$ & $1.97\cdot10^{5}$ & S \\ 
 1.1884 & $13.26\pm0.19$ & $\ \ 93\pm42$ & $5.28\cdot10^{5}$ & B \\ 
 1.2665 & $13.62\pm0.05$ & $\ \ 58\pm5$ & $2.03\cdot10^{5}$  &  \\ 
 1.3206 & $13.11\pm0.10$ & $\ \ 44\pm10$ & $1.16\cdot10^{5}$ & SG \\ 
 1.3395 & $13.85\pm0.03$ & $\ \ 53\pm4$ & $1.70\cdot10^{5}$  & BG \\ 
 1.5377 & $13.41\pm0.05$ & $\ \ 41\pm5$ & $1.04\cdot10^{5}$ &  \\ 
 1.5390 & $13.83\pm0.02$ & $\ \ 54\pm3$ & $1.74\cdot10^{5}$ &  \\ 
 1.5454 & $13.01\pm0.26$ & $108\pm46$ & $7.13\cdot10^{5}$ & S \\ 
 1.6331 & $13.13\pm0.10$ & $145\pm22$ & $1.28\cdot10^{6}$ & G \\ 
 1.6501 & $13.29\pm0.17$ & $239\pm53$ & $3.45\cdot10^{6}$ & GM \\ 
 1.6547 & $13.19\pm0.19$ & $177\pm37$ & $1.89\cdot10^{6}$ & G \\ 
 1.6830 & $12.43\pm0.09$ & $\ \ 47\pm9$ & $1.34\cdot10^{5}$&  \\ \midrule
\multicolumn{5}{c}{HE 0141-3932}\\ \midrule
 1.5945 & $12.75\pm0.05$ & $\ \ 42\pm5$ & $1.06\cdot10^{5}$ &   \\ 
 1.6000 & $12.77\pm0.06$ & $\ \ 53\pm6$ & $1.70\cdot10^{5}$ & S \\ 
 1.6168 & $12.61\pm0.14$ & $\ \ 64\pm15$ & $2.51\cdot10^{5}$ & S \\ 
 1.6195 & $12.76\pm0.05$ & $\ \ 48\pm6$ & $1.40\cdot10^{5}$ & SB \\ 
 1.6285 & $12.80\pm0.07$ & $\ \ 73\pm11$ & $3.20\cdot10^{5}$ & BG \\ 
 1.7176 & $12.87\pm0.03$ & $\ \ 54\pm4$ & $1.75\cdot10^{5}$ & G \\ 
 1.7416 & $13.03\pm0.03$ & $\ \ 67\pm4$ & $2.76\cdot10^{5}$ & B \\ \midrule
\multicolumn{5}{c}{HE 0429-4901}\\ \midrule
 1.8607 & $12.99\pm0.15$ & $\ \ 59\pm17$ & $2.12\cdot10^{5}$ & B \\ 
 1.8642 & $13.10\pm0.04$ & $\ \ 42\pm4$ & $1.06\cdot10^{5}$ &  \\ \midrule
\multicolumn{5}{c}{HE 2225-2258}\\ \midrule
 1.6926 & $13.12\pm0.11$ & $\ \ 77\pm13$ & $3.56\cdot10^{5}$ & B \\ 
 1.7608 & $12.54\pm0.13$ & $\ \ 50\pm13$ & $1.51\cdot10^{5}$ & B \\ 
 1.8031 & $12.35\pm0.15$ & $\ \ 53\pm17$ & $1.69\cdot10^{5}$ & G \\ \midrule
\multicolumn{5}{c}{PG 1634+706}\\ \midrule
 0.8876 & $12.97\pm0.09$ & $\ \ 48\pm11$ & $1.38\cdot10^{5}$ & M \\  
 0.9249 & $13.02\pm0.07$ & $\ \ 53\pm8$ & $1.69\cdot10^{5}$ & SG \\   
 0.9852 & $12.75\pm0.19$ & $\ \ 67\pm26$ & $2.73\cdot10^{5}$ & GM \\ 
 1.1618 & $13.86\pm0.02$ & $\ \ 43\pm1$ & $1.14\cdot10^{5}$ &  \\ 
 1.2038 & $13.00\pm0.06$ & $\ \ 50\pm7$ & $1.51\cdot10^{5}$ & SG \\ 
 1.2255 & $13.69\pm0.02$ & $\ \ 46\pm3$ & $1.27\cdot10^{5}$ &  \\  
 1.2269 & $13.81\pm0.02$ & $\ \ 80\pm4$ & $3.83\cdot10^{5}$ &  \\  
 1.2838 & $13.39\pm0.06$ & $\ \ 41\pm5$ & $1.04\cdot10^{5}$ &  B \\ 
 1.2846 & $13.31\pm0.07$ & $\ \ 60\pm11$ & $2.17\cdot10^{5}$ & B \\ \bottomrule
\end{tabular}
\end{minipage}
\end{table}

The comoving path length was determined using Eq.\ \ref{eq:path}. Just parts of the spectrum with a clearly defined continuum, that are thus applicable for WHIM detection, were included. The resulting total blocking corrected path length has a value of $\Delta X=1.964$. Under the assumption of thermal broadening, the ionisation fraction was calculated with Eq.\ \ref{eq:ion} for each absorber. The baryon density was then estimated with Eq.\ \ref{eq:density}.

The resulting density parameter is $\Omega_{\rm{b}}(\rm{BLA})=(0.91\pm0.05)\cdot10^{-3}$ for the good sample and $\Omega_{\rm{b}}(\rm{BLA})_{\rm{tent}}=(8.4\pm1)\cdot10^{-3}$ for the whole sample, including the tentative detections. In the lower redshift range $0.9\leq z<1.5$ a density parameter of $\Omega_{\rm{b}}(\rm{BLA})=(1.50\pm0.07)\cdot10^{-3}$ $\left(\Omega_{\rm{b}}(\rm{BLA})_{\rm{tent}}=(4.8\pm0.6)\cdot10^{-3}\right)$ was found. The higher redshift interval $1.5\leq z\leq1.9$ yielded $\Omega_{\rm{b}}(\rm{BLA})=(0.33\pm0.02)\cdot10^{-3}$ $\left(\Omega_{\rm{b}}(\rm{BLA})_{\rm{tent}}=(12\pm1)\cdot10^{-3}\right)$. Applying the strict detection limit of $\log\frac{N}{b}\geq11.1$ resulted for the low redshift range in a baryon density of $\Omega_{\rm{b}}(\rm{BLA})_{\rm{tent}}=(3.4\pm0.4)\cdot10^{-3}$ and for the high redshift range in $\Omega_{\rm{b}}(\rm{BLA})_{\rm{tent}}=(0.68\pm0.07)\cdot10^{-3}$, using the whole sample.

To consider effects of photoionisation the modified ionisation fraction (Eq.\ \ref{eq:ion2}) was used. Thus a baryon density of $\Omega_{\rm{b}}(\rm{BLA})=(2.2\pm0.1)\cdot10^{-3}$ $\left(\Omega_{\rm{b}}(\rm{BLA})_{\rm{tent}}=(14\pm2)\cdot10^{-3}\right)$ was found.

\begin{figure}
\label{fig:lines}
\centering
\includegraphics[bb=40 498 299 790]{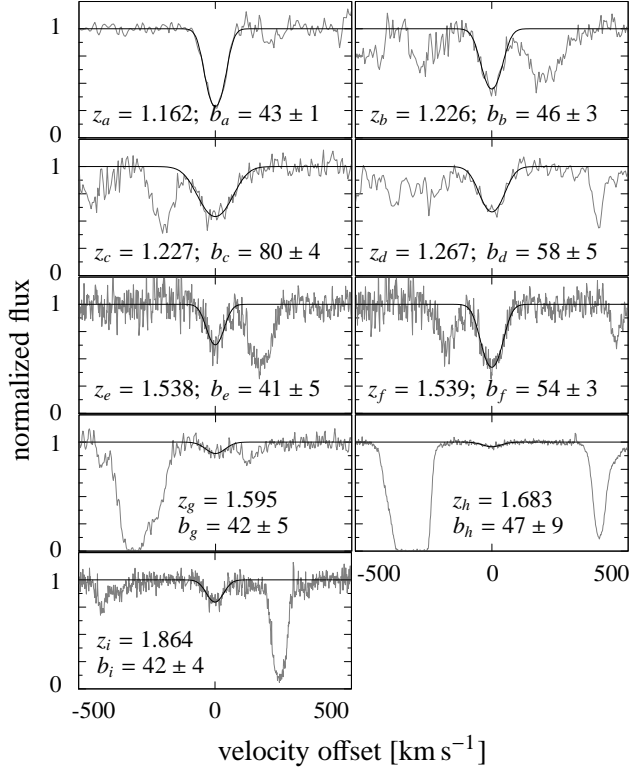}
\caption{Plots of the most reliable WHIM detections sorted by redshift. The Doppler parameter of each absorber is given in $\unit{km\,s^{-1}}$. HE 0515-4414: d, e, f, h. HE 0141-3932: g. HE 0429-4901: i. PG 1634+706: a, b, c.}
\end{figure}

\section{Discussion}
\label{sect:discussion}
The density parameter of the total baryonic matter, determined from the nucleosynthesis model and measurements of primordial element abundances, is expected to be $\Omega_{\rm{BM}}\approx0.04$, using a Hubble constant of $H_{0}=\unit[72]{km\,s^{-1}}$. A baryon density of $\Omega_{\rm b}({\rm Ly}\alpha)=0.010\pm0.001\cdot T_4^{0.56}$ for the diffuse photoionised IGM thus represents a fraction of at least $\sim25\%$ to the total baryons for temperatures $T>\unit[10^4]{K}$. The WHIM was found to contribute with at least $\sim2.3\%$ (complete sample: $22\%$) to the total baryonic matter. Using the modified ionisation fraction (Eq.\ \ref{eq:ion2}) yielded a contribution of leastwise $\sim5.5\%$ to the total baryons.

While the complete sample should overestimate the baryon density, the good sample is expected to underestimate the baryon content significantly and can thus be seen as a lower limit. Many absorbers that might trace the WHIM were excluded, since non-thermal broadening may play a role. Very broad absorbers with Doppler parameters $b\gtrsim\unit[100]{km\,s^{-1}}$, corresponding to temperatures of $T\gtrsim\unit[6\cdot10^5]{K}$, were all marked as tentative since the line shape is not discernable with the current data quality available. Smallest fluctuations in the quasar continuum make a determination of the line shape impossible. Since these very broad and thus hot absorbers have a high ionisation fraction, they might inhabit large amounts of baryons that remain undetected. Even with higher resolution data with a better $S/N$ it could turn out to be impossible to trace gas at temperatures $T\gtrsim\unit[10^6]{K}$ with this method since the quasar continuum is not completely flat and its formation is largely unknown. Effects of non-thermal broadening and continuum fluctuations would not be distinguishable.

Hydrodynamical simulations are able to reproduce the features of the Lyman $\alpha$ forest quite well and are thus seen as a reliable tool for studying the IGM. \citet{Dave2001} found that at $z\approx1.5$ about $70\%-80\%$ of the baryons reside within the diffuse photoionised IGM, which is  higher than our result. The reason can be found in the uncertain intensity of the ionising background radiation, the unknown temperature and a possibly incomplete line sample. The WHIM is supposed to contain $10\%-20\%$ of the total baryons. As suspected our result lies beneath that value.

The result indicates that the IGM is in a stage of evolution at $z\approx1.5$. While for temperatures $T>\unit[10^4]{K}$ at $1.5\leq z<1.9$ still over $30\%$ of the baryons reside in the diffuse photoionised IGM, the value shrinks to $20\%$ for $0.9\leq z\leq1.5$. The obvious increase in the baryon density of the warm-hot phase of the IGM from $\sim0.8\%$ for the high redshift sample to $\sim4\%$ for the low redshift sample shows that indeed a considerable fraction of the baryons has been shock heated to higher temperatures at the analysed redshift range. However, the result has to be treated with care since the two redshift intervalls were traced with different instruments. the atmospheric cut-off at $\lambda_{\rm{cut-off}}\approx\unit[3000]{\AA}$ corresponds to a redshift of $z_{\rm{cut-off}}\approx1.5$ for hydrogen absorption. Thus the low redshift data were taken with STIS, while the higher redshift data were obtained with UVES. However, since there are more regions with a high $S/N$ in the UVES data then in the STIS data, more very broad BLAs that contain high amounts of baryons would be detectable in the high redshift range. Thus the general trend is not suspected to be created by the change of instruments. Though a lower $S/N$ could result in a higher number of falsly as broad features interpreted line blends it also increases the number of lines that are neglected because the shape could not be determined satisfactorily. Additionally a systematic effect in the change of instruments would have the same influence on the narrow as on the broad lines, which is not supported by the observations. We found WHIM absorbers up to a redshift of $z=1.863$. Using the strict detection limit of $\log\frac{N}{b}>11.1$, the complete sample also shows a considerable increase of baryons with lower redshift. The reason for the opposite behaviour of the complete sample for the general detection limit lies again in the different quality of the UVES and STIS data. Very broad absorbers, which contain the highest amounts of baryons, are all marked as tentative and are just detected in the highest quality regions of the UVES data. They would not be detectable in the STIS data, and thus at lower redshifts.

It is evident that the WHIM contributes to an important degree to the total baryon density, not only in the local universe but also at higher redshifts $1\lesssim z\lesssim2$. A study of the detailed evolution of the WHIM, to solve the missing baryon problem once and for all, is of great interest. The future awaits several X-ray missions that should be able to detect diffuse emission from WHIM sources. Until now the search for broad Lyman $\alpha$ absorption seems to be the most promising approach. High resolution spectra with a good $S/N$ would be required to make a reliable estimation of the baryon content of the WHIM at least for the temperature range $T\sim\unit[10^5-10^6]{K}$. The Cosmic Origins Spectrograph (COS), a new instrument to be installed on the Hubble Space Telescope during the 2008 servicing mission, will extend the current STIS data for the low redshift range $z\leq1.5$.
\bibliographystyle{aa}
\bibliography{7283}
\end{document}